# Valuation of Crypto-Currency Mining Operations

J. Berengueres [*†]

**Abstract.** Traditionally, the Net Present Value method is used to compare diverging investment strategies. However, valuating crypto-projects with fiat-based currency is confusing due to extreme coin appreciation rates as compared to fiat interest rates. Here, we provide a net present value method based on using crypto-coin as the underlying asset. Using this method, we compare HODL vs. mining, we also provide a sensitivity analysis of profitability.

1. Net Present Value.   2. GPU mining.   3. Valuation.

## 1. Introduction

In the current crypto mining boom[1,2], two opposed views exist on profitability of mining operations. One view states that mining is profitable, the other states that HODLing the coin is more profitable[3,4]. However, given any economic criterion[5] there is only one optimal strategy. A widely-used criterion to compare investments so far has been Net Present Value (NVP)[6,7]. In the crypto case, we can use it to aggregate the future cash flows that a miner will produce during its lifetime. However, the NPV method is not straight forward to interpret because it depends on the interest rate of the fiat money chosen to measure the cash flow. For example, given a miner that produces coins, assessing its NPV by aggregating future discounted cash flows at a given interest (aka discount rate) is complex because it is not clear what interest rate should be applied. It is also hard to account or estimate the effect of the appreciation of the coin. Moreover, (unlike fiat) the mined asset, does not depreciate, quite the opposite. This poses questions on whether it is appropriate to discount cash flows (coins) that are basically not inflationary. An alternative, is to use the Net Coin Value (NCV). The NCV is the sum of the coin flow that a mining operation will produce minus all the operating expenses (not capital) valued at the price of the coin on the day of the purchase of the equipment.

$$NCV = \sum_{i=1}^{n} C_i \qquad (1)$$

where $C_i$ is the amount of mined coin at the end of one day (24h) minus electricity bill, $n$ is the last day of mining, and $C_i$ is defined as,

$$C_i = (1-k)\frac{M_0}{(1+r)^i} - \frac{e}{P_0} \qquad (2)$$

---

[†] J. Berengueres (jose@uaeu.ac.ae) is Asc. Professor of Computer Science at CIT, UAE University, UAE.

where *k* represents various fees (pool fee + mining software fee + hosting and admin overheads), $M_0$ is amount of the coin mined on day 0, and *r* is the daily growth of hashing capacity of all miners mining the coin, *e* is the daily electricity bill divided by the price P of the coin on day of purchase of the equipment (i=0). From this, it follows that the payback time happens on the first day of mining that verifies

$$NCV(i) > \frac{Cost\ miner}{P_0} \qquad (3)$$

The time to double the initial investment is then the day *i* that verifies:

$$NCV(i) > 2\frac{Cost\ miner}{P_0} \qquad (4)$$

## 2. Valuation Examples

### A. GPU Mining case

To illustrate NCV, let's use a real example based on a rig composed of 8 x GPU RX580 and the Claymore mining software. In this case, an investor would be interested in finding out whether to invest in the rig or to HODL coins. Fig. 1 shows a daily cash flow for a scenario where electricity costs 0.19EUR/kwh (Amsterdam rate); the rig costs $6,756 of which approximately $4,000 is the cost of the GPUs and the rest belongs to Power Supply Units and motherboard, etc... A 10% admin fee on the mined coins is levied to account for pool fees (1%), Claymore miner fee (1%), rig hosting fee (typically 5 to 25%). In the chart, Y is coin and X is days. Four cash flows lines are shown: line 1: HODL is spending the same amount the rig costs into buying coins and holding them, the curve 2 is the cash flow corresponding to buying a rig on day 0 with coins @$P_0$ and then accruing the subsequent coins produced. Coin production declines as more mining power is added to the Ethereum network. We use a decline estimation based on exponential growth of the hashing capacity at rate *r* (Eq. 2), curve 3 shows the NCV for the same rig, but assuming linear network capacity growth that corresponds to a linear interpolation of the past 12 months provided by coinwarz[8]. In the case of the exponential growth, we assume a 0.45% daily growth rate (same as the BTC network during some periods in the past, as a fastest-case scenario). Finally, line 4 shows the daily coin flows if the network hash capacity was to grow at the same rate as Moore's Law, the most optimistic scenario for miners. This is the best-case scenario with the slowest decline rate.

As we can see from Fig. 1, the rig recovers the initial investment fast at the beginning and slower later. However, at current estimated network growth rate, it never recoups the cost when we measure value in NCV. Then, about a year since operation start, the rig will cost more to operate than what the electricity costs. The accumulated coin produced by the rig never surpasses HODL. We assume price of coin constant, and this assumption overestimates the electricity cost measured in coins if the coin appreciates.

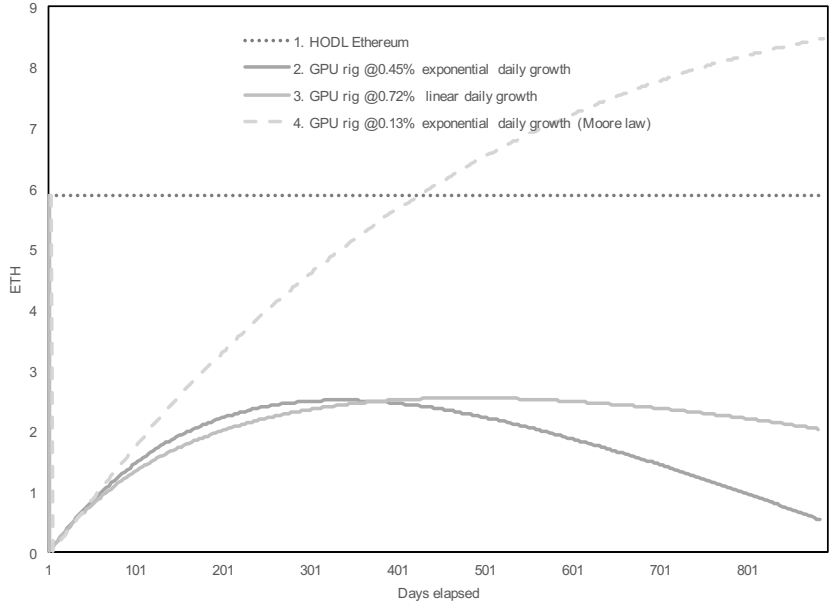

Fig. 1. Fig.1 Accumulated daily cash flow for four investment scenarios at EUR 0.19kwh (Amsterdam prices). At estimated global hashing capacity growth, the max NCV for mining occurs after 1 year and underperforms HODL by more than 50%.

B. Bitcoin Cash Mining case

In this example, we will address the profitability of an S9 Miner with parameters as per Table 1. Fig. 2 shows the evolution of NCV measured in Bitcoin Cash (BCH). It compares the NCV for a miner delivered on payment day versus a miner delivered after 4 or 5 months of prepayment. Table 2, summarizes the dramatic effect that delays in delivery of S9 mining machines have on profitability.

In Fig. 3 we can also see that the NCV @ free electricity provides a hard cap on how much value a rig can produce. We can also see that network growth, rather than electricity cost, is the driving factor impacting the NCV of a mining operation. For example, halving the electricity cost from 0.19 to 0.10 will only increase the (max) NCV from 2.5 to 3.5 coins. Table 3 offers a qualitative sensitivity analysis of impact on profitability. From it we see that delays in delivery and price of the mining equipment are far more important than the daily rate of difficulty increase or the price of the electricity (given typical ranges). In other words, if an equipment is purchased in coin and in advance, the delivery time has an important impact in the total coins mined because it shortens the useful life of the mining equipment exactly when it was most productive: at the beginning. From Table 3 we can see that a mere 140 days of delivery delay results in a loss of 1.6 coins, or more than half of the potential max NCV, as compared to a machine that starts mining immediately after payment.

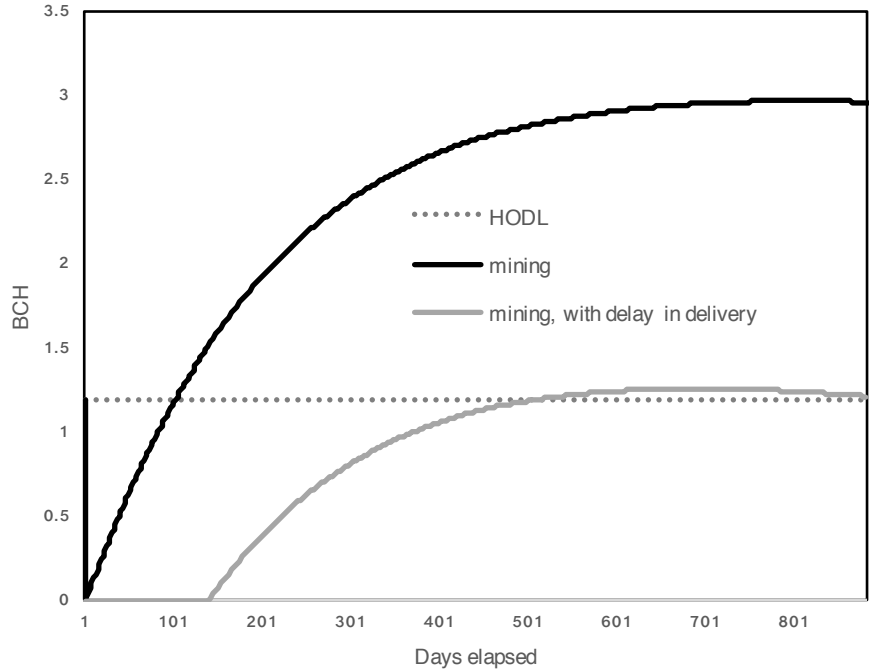

Fig. 2. Prepaying for machines with delays in delivery times has significant impact on profitability at hashing capacity growth levels.

Table 1. Bitmain Antminer S9 parameters

| Item | value |
| --- | --- |
| Price S9 in coins | 1.19149 |
| Price S9 in $ | $2800.00 |
| Difficulty increase daily (exponential) | 0.00450 |
| BCH mined per day per miner* | 0.01702 |
| Cost kwh EUR | 0.03000 |
| Electricity cost in BCH / day | 0.00045 |
| kw per miner | 1.6 |
| Admin fee | 0.10000 |
| BCH price at miner purchase time | $2350.00 |
| Max NCV | BCH 2.92 |

*Own estimate

Table 2. Effect of Delay on Profitability

| Item | No delay in delivery | 140 days delay |
| --- | --- | --- |
| Max NCV in BCH | 2.925 | 1.253 |
| Max NCV in BCH @$P_0$ | $6875 | $2946 |
| Max ROI | 2.45 | 1.05 |

Table 3. Sensitivity analysis ROI

| Rank of top factors | Sensitivity[*] |
|---|---|
| 1st Delay to start to mining from purchase date | s>1 |
| 2nd Cost of rig | s>1 |
| 3rd Network growth rate in % daily | s~1 |
| 4th Electricity price (impact decreases if coin appreciates) | s~1 |

*s~ 1 means proportional*

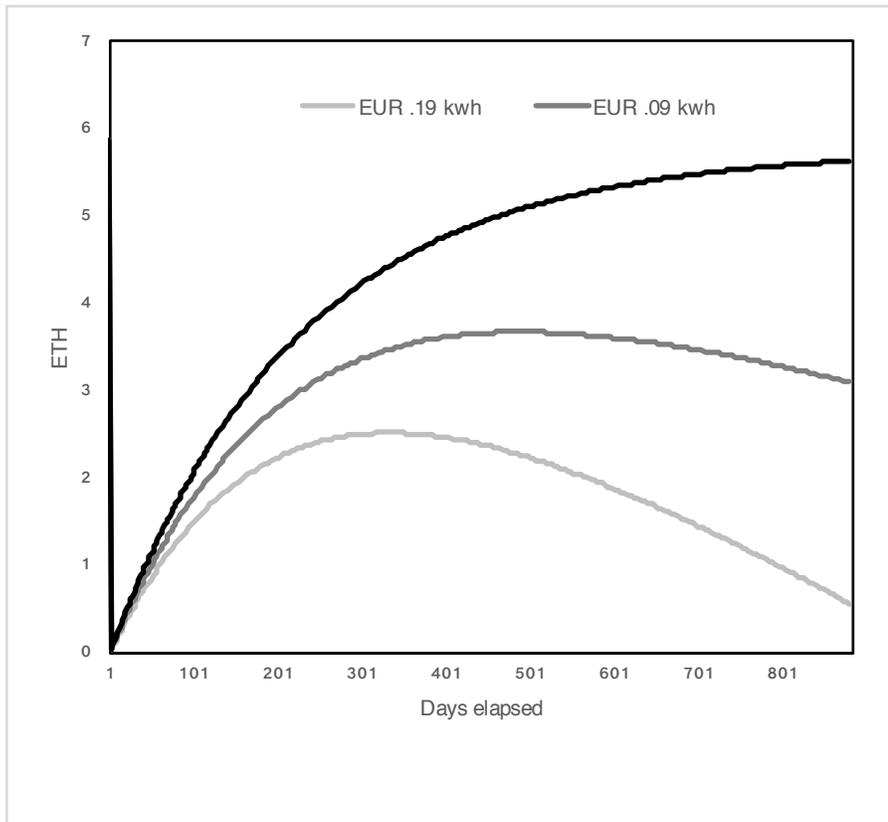

Fig. 3. Effect of different electricity prices in the life span of the miner. Dark line top is for free electricity. Doubling the electricity price from 0.09 to 0.19 does not impact NCV proportionately.

C. Benefits of using NCV to evaluate projects

Here we will compare NPV vs. an NCV. We show how using NPV can lead to suboptimal investment decisions. Let's assume the case in Fig. 1, a GPU rig to mine Ethereum. Clearly the NCV value of the rig is less than the cost of the machine in coins at the time of purchase (negative ROI). However, if Ethereum was to triple in price since the purchase of the rig (as it happened), and the investor measures the cash flows in USD rather than in ETH he could be fooled into believing that the rig was a good investment decision because the value of the total mined coin after a few months was higher than the cost of the rig in USD, and therefore ROI was positive, not realizing that HODL would have been twice as profitable in fiat terms.

## 3. Other factors that impact profitability

### A. Cooling cost

Another factor often overlooked in mining farm projects are the costs of AC, the fire insurance, and so on. For example, in countries, such as Germany, a mining license is required to mine Ethereum even at one's home. In other latitudes, cooling is a significant challenge in hot weather places such as Dubai. In summer, outdoor temperatures can reach up to 55C and cards must not operate at high temperatures. This cost cannot be overlooked in a profitability analysis. Conveniently, AC and heat pumps have Coefficients of Performance (CoP). These cannot thermodynamically exceed 2 to 4 depending on the design. This means that for a CoP =2 for every 2 kw of GPU heat that we want to dissipate, at least 1kw of power must be used by the heat pump or AC just to maintain the temperature inside the mining farm stable. Therefore, Eq. 2 becomes:

$$C_i = (1 - k) \frac{M_0}{(1 + r)^i} - (1 + \frac{1}{CoP}) \frac{e}{P_0} \qquad (5)$$

Where a *CoP* value is typically 2 to 4, and the cost of the AC equipment would be added to rig cost on a pro rata basis.

### B. Mining with cards without warranty

Overclocking cards increases hashing power between 10 to 20%. For example, from 27MHz/s to 32MHz/s in the case of a AMD Rx580 (Ethereum case). This is not depreciable. On the other hand, because mining 24/7 and overclocking abuses the hardware card manufacturers do not issue warranty on so-called mining-cards such as the NVidia p106. Therefore, many farms prefer to use commercial 2-year warranty GPU cards such as the NVidia 1060-70-80 consumer series. While the benefits of overclocking with warranty are substantial, overclocking increases the rate of failure of cards the warranty compensates for that.

### C. Cycle life time

The NCV peak provides an estimate of when a card becomes unprofitable to operate. Assuming all else constant, we see effective lifetimes of 18 months. Therefore, cost of the card it should be treated as fungible cost, not a capital expenditure in NPV calculations. The 18-month lifetime is surprisingly accurate. For example, today, mining Ethereum with an NVidia GeForce 1060 (launched to market on May 2016, hashing power 10Mh/s), does not produce even 1/3rd of coin produced by an NVidia 1070 card (27Mh/s). It produces close to zero due to the way mining pools work (timeouts). If from experience, we consider that the card value for mining drops to zero in 18 months (n=540 days) and consider it as a fungible (not a capital expense) then the marginal cost of mining verifies:

$$C_i = \frac{P_{CARD}}{n} \qquad (6)$$

Where $C_i$ is Eq. 5, $P_{CARD}$ is the price of the GPU card in coin at purchase time. Then the number of coin used to compare with HODL alternative would include all CAPEX in mother boards, PSU, AC and cabling and exclude the cost of GPU cards. Eq. 6 is appropriate because $P_{CARD}$ is correlated with the price of coins that the card can mine at purchase time while the rest of the equipment is not, and because the life time of the rest of the rig is greater than 18 months. From Eq. 6 we can now estimate the marginal pairs (prices of cards, electricity) that make mining marginally profitable. As coin returns diminish, and substituting i=n, we can now also forecast if a card will reach its end of lifetime due to obsolescence (Moore's Law boundary) or because a high price of electricity, in which case *n* should be shortened accordingly in Eq. 6

## 4. Conclusions

We have shown how to use Net Coin Value method to value mining operations using Ethereum and Bitcoin Cash as the underlying asset. This method, offers a simpler alternative to the discounted cash flows method which is not suited for underlying assets that do not depreciate in time. From a qualitative sensitivity analysis, we conclude that there are four main factors that impact profitability in NCV terms. Delivery delay (the time from pre-pay to switch on) has a disproportionate effect on the NCV. Hence, for mining equipment sellers, the easiest way to adjust demand might not be altering the price tag but altering the delivery date on pre-orders. Finally, we can now address the miner's profitability paradox: Mining seems never profitable for new entrants because existing miners that can simply upgrade GPU in their data centers have an unfair capital advantage compared to new entrants, who must invest in the surrounding infrastructure such as cooling, cabling and admin personnel from scratch. We hope this analysis helps to clarify profitability analysis of mining farms.

## Acknowledgement

Thanks to Frank Webber, Beyumi K., Carlos Domingo, Ferran Pujol and Eneko Knorr for fruitful discussion and feedback.

## Author Contributions

J. Berengueres wrote 100% of the text.

## Conflict of Interest

The author is partner at ai2co.com

## Notes and References

**8**